\documentclass[twocolumn,           
               showpacs,            
               nopreprintnumbers,     
               aps,                 
               prd,          	    
               a4paper,             
               groupedaddress,  
               unsortedaddress,
               nofootinbib,         
               tightenlines,        
               floats               
               ]{revtex4}

\usepackage{graphicx}
\usepackage{amssymb,latexsym}
\usepackage[draft=false]{hyperref}

\begin{document}




\title{Braneworld flow equations}
\author{Erandy Ram\'{\i}rez and Andrew R.~Liddle}
\affiliation{Astronomy Centre, University of Sussex, 
             Brighton BN1 9QH, United 
Kingdom}
\date{\today} 
\pacs{98.80.Cq \hfill astro-ph/0412556}
\preprint{astro-ph/0412556}


\begin{abstract}
We generalize the flow equations approach to inflationary model building to the 
Randall--Sundrum Type II braneworld scenario. As the flow equations are quite 
insensitive to the expansion dynamics, we find results similar to, though not 
identical to, those found in the standard cosmology.
\end{abstract}

\maketitle

\section{Introduction}

The flow equations approach, pioneered by Hoffman and Turner \cite{HT} and by 
Kinney \cite{K} (see also Refs.~\cite{flowpapers,pwmap}), is a means of 
generating 
large numbers of inflation models via a random process, permitting for example 
that ensemble to be compared with observational results as done by the Wilkinson 
Microwave Anisotropy Probe team \cite{pwmap}. It has however been pointed 
out by Liddle \cite{flow} that the flow equations algorithm is actually rather 
insensitive to the equations governing inflationary dynamics. In particular, the 
Friedmann equation is not needed to determine the flow equation trajectories in 
the space of slow-roll parameters; its only role is to measure the amount of 
expansion taking place along those trajectories.

In this {\em Brief Report}, we make an explicit study of the effect of modifying 
the Friedmann equation by considering the flow equations in the simplest 
braneworld inflation scenario, based on the Randall--Sundrum Type II model 
\cite{RSII}.

\section{Standard cosmology flow equations}

The flow equations are a set of differential 
equations linking a set of slow-roll parameters defined from the Hubble 
parameter $H$. In the standard cosmology, following the notation of Kinney 
\cite{K}, the parameters can be defined as
\begin{eqnarray}
\label{e:eps}
\epsilon(\phi) & \equiv &  \frac{m_{{\rm Pl}}^2}{4\pi} \left( 
\frac{H'(\phi)}{H(\phi)} \right)^2 \,; \\
\label{e:srdef}
^{\ell}\lambda_{{\rm H}} & \equiv & \left( \frac{m_{{\rm Pl}}^2}{4\pi}
	\right)^\ell \, \frac{(H')^{\ell-1}}{H^\ell} \, 
	\frac{d^{(\ell +1)} H}{d\phi^{(\ell +1)}} \quad ; \quad \ell \ge 1 \,,
\end{eqnarray}
where primes are derivatives with respect to the scalar field $\phi$.  Using 
the relation
\begin{equation}
\frac{d}{dN} = \frac{m_{{\rm Pl}}^2}{4\pi} \, \frac{H'}{H} \, 
\frac{d}{d\phi} \,,
\end{equation}
where we define the number of $e$-foldings $N$ as {\em decreasing} with 
increasing time, yields the flow equations
\begin{eqnarray}
\label{e:flow}
\frac{d\epsilon}{dN} & = & \epsilon(\sigma+2\epsilon) \,; \nonumber \\
\frac{d\sigma}{dN} & = & -5\epsilon\sigma - 12 \epsilon^2 +
	2(^2\lambda_{{\rm H}}) \,; \\
\frac{d(^\ell\lambda_{{\rm H}})}{dN} & = & \left[\frac{\ell-1}{2} \, \sigma
	+(\ell-2)\epsilon \right](^\ell\lambda_{{\rm H}})+^{\ell+1}\!\!\!
	\lambda_{{\rm H}} \; ; \; \ell \ge2 \,, \nonumber
\end{eqnarray}
where $\sigma \equiv 2(^1\lambda_{{\rm H}})-4\epsilon$ is a convenient 
definition. 
 
As pointed out in Ref.~\cite{flow}, these equations actually have limited 
dynamical input from inflation, since the above derivation has been made without 
reference to the Friedmann equation. Indeed, 
if written in the form $d/d\phi$ they are a set of 
identities true for any function $H(\phi)$, and the reparametrization
to $d/dN$ modifies only the measure along the trajectories, not the 
trajectories themselves. In that light it seems surprising that they 
can say much about inflation at all, but it turns out that the flow 
equations can be viewed as a (rather complicated) algorithm for 
generating functions $\epsilon(\phi)$ which have a suitable form to be 
interpreted as inflationary models \cite{flow}. 

Ref.~\cite{flow} implied that the flow equation predictions ought to be little 
changed by moving to the braneworld, although this modifies the Friedmann 
equation. However this statement needs explicit 
justification, because although the trajectories are unaffected, there are 
several changes which affect the predictions: the measure of $e$-foldings along 
the trajectories changes, the 
endpoints of the trajectories change, and the equations relating the slow-roll 
parameters to the observables change. In this short article we investigate these 
effects, restricting ourselves to the high-energy regime of the Randall--Sundrum 
Type II model \cite{RSII}.

\section{Braneworld flow equations}

We follow the notation of Ref.~\cite{RL}.
In the Randall--Sundrum Type II braneworld model \cite{RSII}, the Friedmann 
equation in the 
high-energy regime can be written as
\begin{equation}
H=\frac{4\pi}{3M_5^3}\rho \,,
\label{ef:bw}
\end{equation}
where $M_{5}^3$ is the five-dimensional Planck mass, related to the brane 
tension $\lambda$ by
$M_5^3\equiv(4\pi\lambda/3)^{1/2} m_{{\rm Pl}}$.
The scalar wave equation is unchanged, and a useful expression is
\begin{equation}
\dot{\phi}=-\frac{M_5^3}{4\pi}\frac{H'}{H} \,.
\label{eq:phi}
\end{equation}

Although they could be used, the standard cosmology definitions of the slow-roll 
parameters, Eqs.~(\ref{e:eps}) and (\ref{e:srdef}), are actually not very 
convenient in the braneworld scenario, in particular because $\epsilon = 1$ is 
no longer the condition to end inflation. Instead, following Ref.~\cite{RL}, we 
define new slow-roll parameters relevant to the high-energy regime as follows:
\begin{eqnarray}
\epsilon_{\rm H}&=&\frac{M_5^3}{4\pi}\frac{H'^2}{H^3}\,; \\
\eta_{\rm H}+\epsilon_{\rm H}&=&\frac{M_5^3}{4\pi}\frac{H''}{H^2} \,. 
\end{eqnarray}
By analogy
to the standard cosmology case, we define the
higher-order slow-roll parameters as
\begin{equation}
^{\ell}\lambda_{{\rm H}}  \equiv  \left( \frac{M_{ 5}^3}{4\pi}
	\right)^\ell \, \frac{(H')^{\ell-1}}{H^{2\ell}} \, 
	\frac{d^{(\ell +1)} H}{d\phi^{(\ell +1)}} \quad ; \quad \ell
	\ge 1 \,,
\label{e:srbw}
\end{equation}
where one can identify $^1\lambda_{1}\equiv\eta_{\rm H}+\epsilon_{\rm H}$.

{}From Eq.~(\ref{eq:phi}), the relation between $\phi$ and $N$, measuring the 
length along the 
trajectories, is changed to 
\begin{equation}
\frac{d}{dN}=\frac{M_5^3}{4\pi}\frac{H'}{H^2}\frac{d}{d\phi} \,.
\label{e:nphibw}
\end{equation}
{}From these definitions, and taking the convention \mbox{$\dot{\phi}>0$,} we 
find 
a set of flow equations
\begin{eqnarray}
\frac{d\epsilon}{d\phi} & = & \frac{H'}{H}(2\eta_{\rm H}-\epsilon_{\rm
H})\,;  \\
\frac{d\eta_{\rm H}}{d\phi} & = &
\frac{H'}{H}\left[\frac{^2\lambda_{\rm H}}{\epsilon_{\rm
H}}-4\eta_{\rm H}-\epsilon_{\rm H}\right]\,;\\
\frac{d(^{\ell}\lambda_{\rm H})}{d\phi} & = & \frac{1}{\epsilon_{\rm H}}
\frac{H'}{H}\left\{ \left[(\ell-1)\eta_{\rm H}-(\ell+1)\epsilon_{\rm
H}\right](^{\ell}\lambda_{\rm H}) \right. \quad \\
&&  \quad \quad \left. +^{(\ell+1)}\lambda_{\rm H}\right\}\,; \quad
\ell\ge 2 \,, \nonumber
\end{eqnarray}
which leads to the following 
set of flow equations for the braneworld
\begin{eqnarray}
\label{fe:bw}
\frac{d\epsilon_{\rm H}}{dN} & = & \epsilon_{\rm H}(\sigma_{\rm H}+
3\epsilon_{\rm H}) \,; \\
\frac{d\sigma_{\rm H}}{dN} & = & -8\epsilon_{\rm H}\sigma_{\rm H}
-30\epsilon_{\rm H}^2+2(^2\lambda_{\rm H}) \,; \\
\frac{d(^{\ell}\lambda_{\rm H})}{dN} & =&
\left[\frac{\ell-1}{2} \, \sigma_{\rm H}+(\ell-3)\epsilon_{\rm
H}\right](^{\ell}\lambda_{\rm H}) \\
&& \quad \quad +^{(\ell+1)}\lambda_{\rm H}; \,\, \ell\ge 2 \,, \nonumber
\end{eqnarray}
where $\sigma_{\rm H} \equiv 2 \eta_{\rm H} -4\epsilon_{\rm H}$.

Note that the braneworld slow-roll parameters were purposefully defined 
\cite{RL} so that 
the 
end of inflation is at $\epsilon_{\rm H} = 1$, and so that 
standard expression for the spectral index $n$ still applies; however the 
expression for the tensor-to-scalar ratio $r$ is modified from the usual $r 
\simeq \epsilon$.\footnote{The recent literature has two different definitions 
of $r$ in common use, the other one being 16 times this one. We use this 
convention to follow Kinney \cite{K}.} The observable quantities at first-order 
are given by \cite{BWperts}
\begin{eqnarray}
r&\simeq&\frac{3}{2}\epsilon_{\rm H} \,;\\ 
n&\simeq& 1+\sigma_{\rm H}\,;\\ \label{e:rnbw}
\frac{dn}{d{\rm ln} k}&\simeq& 30\epsilon_{\rm H}^2+8\epsilon_{\rm
H}\sigma_{\rm H}-2(^{2}\lambda_{\rm H}) \,.
\end{eqnarray}

Following Ref.~\cite{K}, we can analyze the fixed points 
where all the derivatives vanish, for which the conditions are
\begin{equation}
\epsilon_{\rm H}=^{\ell}\lambda_{\rm H}=0, \quad
\sigma_{\rm H}= {\rm const}\,.
\end{equation}
As in the standard cosmology, these correspond to $r = 0$ and are stable for 
$n>1$. The $n<1$ branch is stable for integration backwards in time.

\begin{figure}
\centering
\includegraphics[width=8cm]{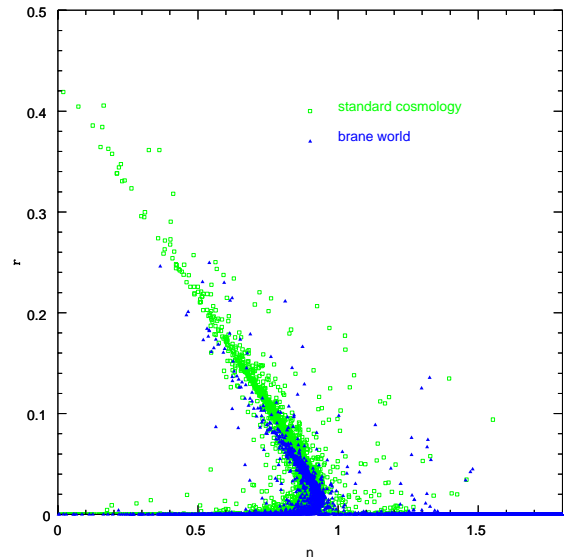}
\caption{Distribution of observables for the standard and braneworld 
cosmologies.}
\label{fig}
\end{figure}

We solve the flow equations following the method of Kinney \cite{K}, both for 
the standard cosmology where we verify his results and for the system of 
braneworld flow equations
written above. We consider 60,000 initial conditions drawn from the ranges 
Kinney uses (taking the same ranges also for the braneworld slow-roll 
parameters). The results are shown in Fig.~\ref{fig}, with observables computed 
to first-order in slow-roll.

For both models, the majority of the points lie effectively on the $r=0$ axis, 
and 
in addition we see in each case the now-familiar swathe of points following a 
tight diagonal locus to large values of $r$ and $1-n$ \cite{HT,K}, plus some 
other scattered points. 
The two distributions are extremely similar, though the swathe for the 
braneworld is slightly below that of the standard cosmology.
We conclude therefore that the change in the
dynamical equations does not significantly affect the distribution of
points in the space of observables.

\section{Conclusions}

We have modified the flow equations approach to implement it in the high-energy 
regime of the Randall--Sundrum Type II braneworld cosmology. Although the flow 
equation trajectories are independent of the dynamical equation driving 
inflation, changes do occur because the measure of length (i.e.~the number of 
$e$-foldings) along the trajectories changes, because the point 
corresponding to the end of inflation changes, and because the formulae giving 
the observables change. Nevertheless, we have shown that 
those effects are small and that the distribution of observables predicted by 
the braneworld flow equations is very similar to that of the standard 
cosmology.

\begin{acknowledgments}
E.R. was supported by Conacyt and A.R.L. by PPARC. We thank Will
Kinney for useful discussions. 

\end{acknowledgments}
 
\end{document}